\documentclass[aip,jcp,preprint,groupedaddress]{revtex4-1}
\usepackage{color}
\usepackage{graphicx, bm, amsfonts, amssymb, amsmath, appendix,setspace,figcaps}
\usepackage{graphicx}
\begin{document}

\title{Density-Dependent Analysis of Nonequilibrium Paths Improves Free Energy Estimates II. A Feynman-Kac Formalism}

\author{David D. L. Minh}
\email[Electronic Address: ]{daveminh@anl.gov}
\affiliation{Biosciences Division, Argonne National Laboratory, Argonne, IL 60439, USA}
\author{Suriyanarayanan Vaikuntanathan}
\email[Electronic Address: ]{svaikunt@umd.edu}
\affiliation{Chemical Physics Program, Institute for Physical Science and Technology, University of Maryland, College Park, Maryland 20742, USA}

\date{\today}

\begin{abstract}
The nonequilibrium fluctuation theorems have paved the way for estimating equilibrium thermodynamic properties, such as free energy differences, using trajectories from driven nonequilibrium processes.  While many statistical estimators may be derived from these identities, some are more efficient than others.  It has recently been suggested that trajectories sampled using a particular time-dependent protocol for perturbing the Hamiltonian may be analyzed with another one.  Choosing an analysis protocol based on the nonequilibrium density was empirically demonstrated to reduce the variance and bias of free energy estimates.  Here, we present an alternate mathematical formalism for protocol postprocessing based on the Feynmac-Kac theorem.  The estimator that results from this formalism is demonstrated on a few low-dimensional model systems.  It is found to have reduced bias compared to both the standard form of Jarzynski's equality and the previous protocol postprocessing formalism.
\end{abstract}

\maketitle

\section{Introduction}
A key goal in computational thermodynamics is the estimation of free
energy differences between equilibrium states.
Challenges in efficiently obtaining accurate values, however, continue
to motivate the development of novel methods.\cite{Chipot2007}
The discovery of new theorems in nonequilibrium statistical mechanics
\cite{Jarzynski1997a, Jarzynski1997b, Crooks1998, Crooks1999,
Crooks2000} have opened up a promising direction: free energy
calculations based on simulations of driven nonequilibrium processes.
\cite{Frenkel2002, Chipot2007}
The most straightforward implementation of this approach involves
performing multiple repetitions of a process
in which a system is driven out of equilibrium by switching an
external parameter $\lambda$
according to a protocol $\Lambda \equiv \lambda(t)$, where $0 \leq t \leq T$.
If the free energy difference of interest is between thermodynamic
states defined by setting $\lambda$ to $A$ and $B$,
then the protocol is defined so that $A$ and $B$ are the end states,
$\lambda(0) \equiv A$ and $\lambda(T) \equiv B$.
The free energy difference between the initial thermodynamic state and
the equilibrium state at any time $t$,
$F_{\Lambda_t} \equiv F({\lambda(t)})-F({\lambda(0)})$,
may then computed by applying,~\cite{Jarzynski1997a, Jarzynski1997b}
\begin{equation}
\label{eq:NEW}
e^{-\beta F_{\Lambda_t}} = \langle e^{-\beta W_t} \rangle_\Lambda
\approx \frac{1}{N} \sum_{n=1}^N e^{-\beta W_t[Z_n]},
\end{equation}
where $\langle\dots\rangle_\Lambda$ is an average over all possible
trajectories (realizations of the process), and $W_t$ denotes the work
done on the system up to time $t$ during a particular trajectory, $Z_n$.
For a finite sample of trajectories $Z_n$ for n = 1, 2, ..., N, the
sample mean provides an estimator for this expectation.

\begin{figure}[tbp]         
\includegraphics[scale=0.4,angle=0]{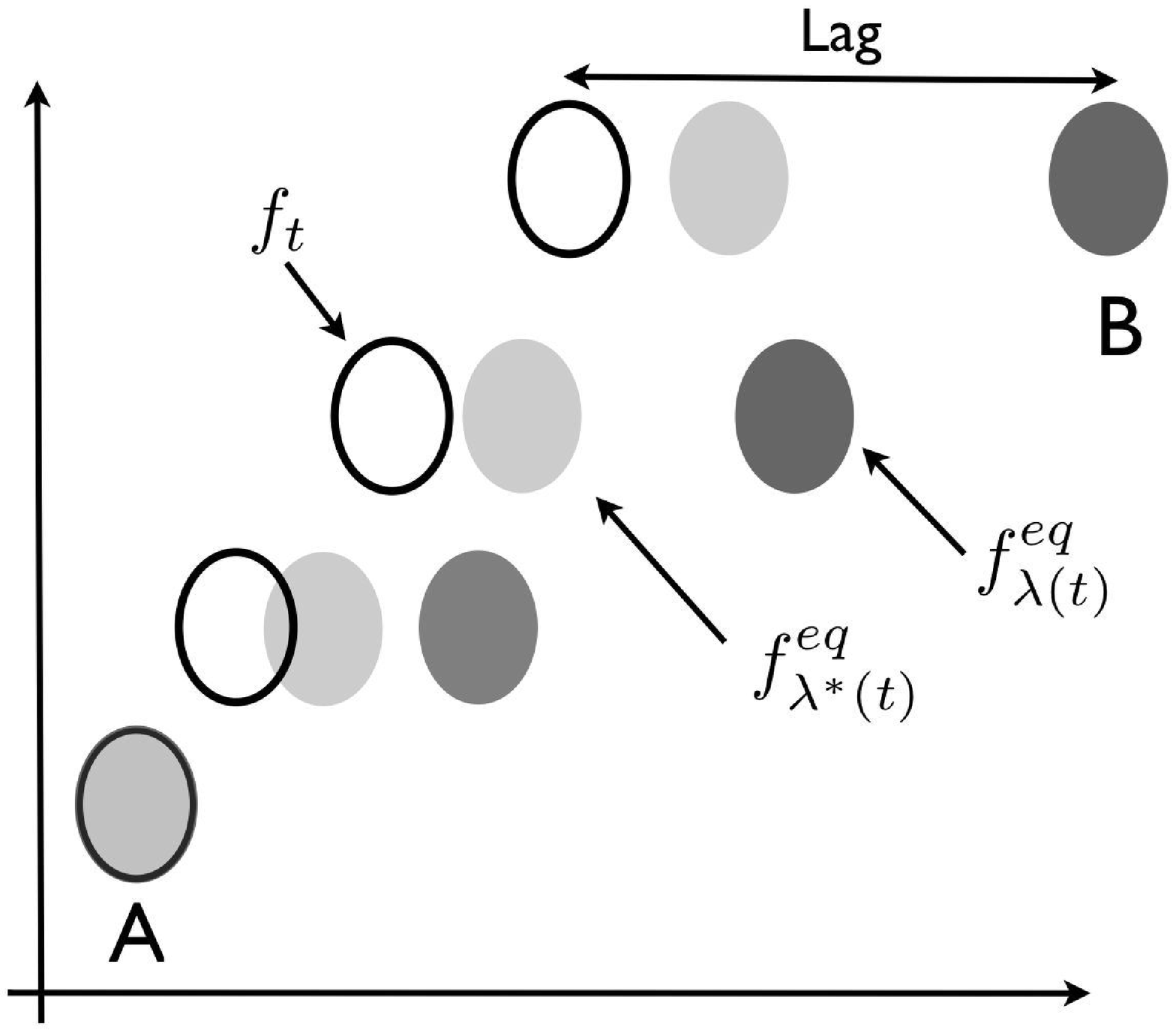} 
\caption{Lag in driven nonequilibrium processes.
Consider a system driven from state $A$ to state $B$ in a finite-time process.
In the above schematic, the ovals represent regions of phase space.
The darkly shaded ovals are regions of phase space 
containing most of the density  $f^{eq}_{\lambda(t)}$ 
of the equilibrium state corresponding 
to the value of the external parameter at time $t$.
The unshaded ovals denote the phase space regions 
containing most of the density $f_t$
actually accessed by the system during the process.
In a reversible process, the two would be indistinguishable.
Since the system is driven out of equilibrium, however, 
a lag builds up between $f_t$ and $f^{eq}_{\lambda(t)}$.
This lag is correlated to dissipation and is ultimately responsible for the poor convergence of free energy estimates based on nonequilibrium processes.
If one is able to obtain a function $\Lambda^* \equiv \lambda^*(t)$ with $\lambda^*(0)=A$ 
such that the equilibrium states $f^{eq}_{\lambda^*(t)}$ are closer to the $f_t$ 
(e.g. the lightly shaded ovals), 
then the convergence of free energy estimates may be improved using Eq. \ref{eq:JarzPP_FK}.
\label{fig:lag_1}}  
\end{figure}

Unfortunately, this estimator often suffers from poor convergence. The expected number of realizations needed to obtain a reliable estimate of $F_{\Lambda_t}$ grows rapidly with the dissipation,
$\langle W_t \rangle_\Lambda - F_{\Lambda_t}$,
that invariably accompanies driven nonequilibrium processes.
\cite{Gore2003, Jarzynski2006, Kofke2006}
In turn, dissipation reflects the lag that develops between the
state of the system
and the equilibrium state corresponding to the instantaneous value of
the external parameter.\cite{Vaikuntanathan2008,Vaikuntanathan2009} 
(See Fig~\ref{fig:lag_1}). This lag is ultimately responsible for the poor convergence of Eq. \ref{eq:NEW}.

The connection between lag and the convergence of the free energy estimator can be better understood by considering two limiting cases. 
First, consider the case of an infinitely slow reversible process.
As the system remains in equilibrium throughout the process, there is no lag.
In this case, convergence only requires a single sample because the work performed along any isothermal quasi-static trajectory, $W_t$, is equal to the free energy difference, $F_{\Lambda_t}$. \cite{LandauLifshitz}
The opposite limit is that of an infinitely fast process, in which Eq. \ref{eq:NEW} reduces to the more familiar free energy perturbation identity. 
Free energy estimates based on this identity converge quickly only if there is significant overlap in the important phase space regions of the end states,\cite{Chipot2007,Kofke2006} 
which in turn reflects the lag. 
Likewise, in the intermediate situation of a finite-time process, the convergence of free energy estimates depends on overlap between the sampled phase space and the important phase space of an equilibrium state of interest.

In order to reduce lag and improve convergence, 
strategies such as 
importance sampling of trajectories~\cite{Athenes2002, Sun2003, Atilgan2004, Ytreberg2004, Athenes2004, Oberhofer2005, Oberhofer2008} 
and ``escorted'' free energy simulations~\cite{Vaikuntanathan2008}
have been introduced.
(For a brief overview, see Ref. \cite{Minh2009}.)
In this paper, we consider an alternate strategy, \emph{protocol postprocessing}.
This strategy involves introducing a function $\Lambda^* \equiv \lambda^*(t)$ 
with $\lambda^*(0) = \lambda(0)$, which we will refer to as the \emph{analysis protocol}.
The central result of this paper (Eq. \ref{eq:JarzPP_FK}) is an expression for the free energy difference $F_{\Lambda_t^*} \equiv F(\lambda^*(t))-F(\lambda(0))$ 
using trajectories generated in the original process
(in which the work parameter is switched according to the protocol $\lambda(t)$). 
While this result is valid for any choice of $\lambda^*(t)$ and reduces to Eq. \ref{eq:NEW} 
for $\lambda^*(t)=\lambda(t)$, 
we will argue that Eq. \ref{eq:JarzPP_FK} provides efficient estimates of the free energy difference $F_{\Lambda_t^*}$ whenever the equilibrium densities corresponding to the analysis protocol $\lambda^*(t)$ have a high degree of overlap with density of the system (See Fig~\ref{fig:lag_1}).

Protocol postprocessing was previous introduced \cite{Minh2009} in the context of importance sampling in path-space. \cite{Athenes2002, Sun2003, Atilgan2004, Ytreberg2004, Athenes2004, Oberhofer2005, Oberhofer2008}
In the present work, we utilize an alternate mathematical formalism,
the Feynman-Kac theorem. \cite{Kac1949}
This formalism is similar to that used in the escorted free energy simulation method, \cite{Vaikuntanathan2008} and indeed, the two methodologies may be used in conjunction with one another.  The new formalism has at least two advantages over the previous method: first, in certain special cases, it is \textit{analytically} a zero-variance estimator.  Secondly, for a few simple model systems, we find that the bias and variance of free energy estimates are substantially reduced.

%
%
\section{\label{sec:FK}Feynman-Kac Formalism}

The derivation of Jarzynski's equality \cite{Jarzynski1997a, Jarzynski1997b} 
from the Feynman-Kac theorem has been well-documented. \cite{Hummer2001a,Hummer2005, Ge2008}
In this section, we recapitulate Hummer and Szabo's derivation \cite{Hummer2001a}  and extend it to protocol postprocessing.

\subsection{Jarzynski's equality}

Consider a classical system with a Hamiltonian, $H_\lambda(z) \equiv H(z;\lambda)$,
that depends on its position in $d$-dimensional phase (or configuration) space, $z$, 
and a parameter vector, $\lambda$.
The system evolves according to dynamics which,
if the temperature $\beta^{-1}$ and $\lambda$ are held constant,
preserve the canonical equilibrium distribution $f^{eq}_{\lambda}(z) \sim e^{-\beta H_\lambda(z)}/Q_\lambda$, where $Q_\lambda = \int dz ~ e^{-\beta H_{\lambda}(z)}$ is a partition function.
These conditions are satisfied by several dynamics such as Hamilton's equations, Langevin dynamics, and the Andersen and Nos\'{e}-Hoover thermostats.

We are interested in driven nonequilibrium processes in which the system is first prepared in equilibrium with $\lambda=\lambda(0)$ and temperature $\beta^{-1}$, after which the external parameters are switched according to the protocol $\Lambda \equiv \lambda(t)$.
Each realization of this process can be described by the trajectory, $Z \equiv z(t)$.
The phase space density $f(z,t)$ of an ensemble of such trajectories evolves according to a Liouville-type equation,
\begin{eqnarray}
\frac{\partial f(z,t)}{\partial t} = \mathcal L_{\lambda(t)} \cdot f(z,t),
\label{eq:evolution}
\end{eqnarray}
As the dynamics preserve the canonical distribution when $\lambda$ is held constant, the operator $\mathcal L_\lambda$ has the property $\mathcal L_\lambda \cdot e^{-\beta H_\lambda(z)} = 0$. \cite{Jarzynski1997b,Hummer2001a}

Hummer and Szabo recognized that the Feynman-Kac theorem provides a solution to the ``sink equation",
\begin{eqnarray}
\frac{\partial g(z,t)}{\partial t} = \mathcal L_{\lambda(t)} \cdot g(z,t) + w(z,t) g(z,t),
\label{eq:mod_evolution}
\end{eqnarray}
where $w(z,t) = - \beta \left( \frac{\partial H_{\lambda(t)}(z)}{\partial t} \right)$, as a path-integral,\cite{Hummer2001a,Hummer2005, Ge2008}
\begin{eqnarray}
g(z,t) = \left< \delta(z - z(t)) e^{\int_0^t ds ~ w(z(s),s)} \right>_{\Lambda}.
\label{eq:FK}
\end{eqnarray}
We remind the reader that the angled brackets $\left< ... \right>_\Lambda$ denote a path-ensemble average, or expectation, over all possible realizations of the described driven nonequilibrium process.

Another solution to Eq. \ref{eq:mod_evolution} is given by 
an improperly normalized Boltzmann distribution,
$ Q_{\lambda(0)}^{-1} e^{-\beta H_{\lambda(t)}(z)}$.
Equating this solution to that from the Feynman-Kac theorem immediately gives,
\begin{eqnarray}
Q^{-1}_{\lambda(0)} e^{-\beta H_{\lambda(t)}(z)} = \left< \delta(z - z(t)) e^{-\beta W_t} \right>_{\Lambda},
\label{eq:weighted_density}
\end{eqnarray}
in which the work $W_t \equiv W_t(Z|\Lambda)$ is defined as,
\begin{eqnarray}
W_t = \int_0^t ds ~ \left( \frac{\partial H_{\lambda(s)}(z(s))}{\partial s} \right).
\label{eq:original_work}
\end{eqnarray}
By integrating both sides over $z$, we obtain 
\begin{eqnarray}
e^{-\beta F_{\Lambda_t} } \equiv \frac{Q_{\lambda(t)}}{Q_{\lambda(0)}}
= \left< e^{-\beta W_t} \right>_{\Lambda}.
\label{eq:Jarz}
\end{eqnarray}

\subsection{\label{sec:PP}Protocol Postprocessing}

In the protocol postprocessing strategy, trajectories are first generated according to the \emph{sampling} protocol $\Lambda \equiv \lambda(t)$.  
Next, a potentially distinct \emph{analysis} protocol $\Lambda^* \equiv \lambda^*(t)$, with $\lambda^*(0)=\lambda(0)$, is introduced.  
This analysis protocol is not used to generate any new trajectories.  
Rather, the previously generated trajectories are used as samples for estimating the free energy difference $F_{\Lambda^*_t} \equiv F({\lambda^*(t)}) - F({\lambda^*(0)})$.  
The standard form of Jarzynski's equality can be seen as a special case where the sampling and analysis protocols are identical.  
While the formalism described below is valid for any $\Lambda^*$, it will not always be advantageous.  
In Section \ref{sec:lag}, however, we will describe how to choose a $\Lambda^*$ that leads to an efficient free energy estimate.

We begin the derivation by formally separating the evolution operator into two terms,
\begin{eqnarray}
\mathcal L_{\lambda(t)} = \mathcal L_{\lambda^*(t)} + \mathcal A(t),
\label{eq:separation}
\end{eqnarray}
where the auxiliary operator $\mathcal A(t)$ represents the difference between the evolution operators given the sampling and analysis protocols.

Now consider a sink equation analogous to Eq. \ref{eq:mod_evolution},
\begin{eqnarray}
\frac{\partial g(z,t)}{\partial t} = \mathcal L_{\lambda(t)} \cdot g(z,t) + w^*(z,t) g(z,t),
\label{eq:mod_evolution2}
\end{eqnarray}
where the function $w^*(z,t)$ not only includes a time-derivative of the Hamiltonian, but also a term with the operator $\mathcal A(t)$,
\begin{eqnarray}
w^*(z,t) = -\beta \left(\frac{\partial H_{\lambda^*(t)}(z)}{\partial t} 
+ \frac{\mathcal A(t) \cdot e^{-\beta H_{\lambda^*(t)}(z)}}{\beta e^{-\beta H_{\lambda^*(t)}(z)}} \right).
\label{eq:mod_sink}
\end{eqnarray}
Here, the operator $\mathcal A(t)$ only acts on the term $e^{-\beta H_{\lambda^*(t)}(z)}$ in the numerator. 
One solution to Eq. \ref{eq:mod_evolution2} is $g(z,t)=Q_{\lambda(0)}^{-1} e^{-\beta H_{\lambda^*(t)}(z)}$.

By equating this solution to the path integral solution obtained from the  Feynman-Kac theorem, we obtain an equation analogous to Eq. \ref{eq:weighted_density}:
\begin{eqnarray}
\frac{ e^{-\beta H_{\lambda^*(t)}(z)}}{Q_{\lambda(0)}} = \left< \delta(z - z(t)) e^{-\beta \mathcal W^*_t} \right>_{\Lambda}.
\label{eq:weighted_density2}
\end{eqnarray}
where the work $\mathcal W_t^* \equiv \mathcal W_t^*(Z|\Lambda^*)$ has the modified form,
\begin{eqnarray}
\mathcal W^*_t = \int_0^t ds ~ \left(\frac{\partial H_{\lambda^*(s)}(z(s))}{\partial s} 
+ \frac{\mathcal A(s) \cdot e^{-\beta H_{\lambda^*(s)}(z(s))}}{\beta e^{-\beta H_{\lambda^*(s)}(z(s))}} \right).
\label{eq:mod_work}
\end{eqnarray}
Integrating over $z$, we obtain a protocol postprocessing form of Jarzynski's equality,
\begin{eqnarray}
e^{-\beta F_{\Lambda^*_t}} = \left< e^{-\beta \mathcal W^*_t} \right>_\Lambda.
\label{eq:JarzPP_FK}
\end{eqnarray}
Again, the angled brackets $\left< ... \right>_\Lambda$ denote a path-ensemble average, or expectation, over all possible realizations of the driven nonequilibrium process with the protocol $\Lambda=\lambda(t)$; the protocol $\Lambda^* = \lambda^*(t)$ has nothing to do with sampling.

To be more concrete, let us consider a system moving with overdamped Langevin (Brownian) dynamics in a one-dimensional potential $U_{\lambda(t)}(z)$. 
The density $f(z,t)$ evolves according to the Smoluchowski equation,
\begin{eqnarray}
\frac{\partial f}{\partial t} = \mathcal L_{\lambda(t)} f =\frac{1}{\zeta} \frac{\partial}{\partial z} \left( U_{\lambda(t)}'(z) f \right) + D \frac{\partial^2}{\partial z^2}f,
\end{eqnarray}
where $D^{-1} = \beta \zeta$ is the diffusion coefficient and the prime symbol represents a derivative with respect to $z$.

Given an analysis protocol $\Lambda^*=\lambda^*(t)$, the auxiliary operator $\mathcal A(t)$ for this example system is defined as,
\begin{equation}
\mathcal A(t) \cdot f \equiv -\beta D \frac{\partial}{\partial z} \left( \Delta U'(z,t) f \right),
\end{equation}
where
\begin{equation}
\Delta U(z,t) \equiv U_{\lambda^*(t)}(z) - U_{\lambda(t)}(z).
\end{equation}
Substituting this expression into Eq. \ref{eq:mod_work}, we obtain a modified form of the work,
\begin{eqnarray}
\mathcal W^*_t 
& = & \int_0^t ds \left( \frac{\partial U_{\lambda^*(s)}(z(s)) }{\partial s} 
	-  \frac{\beta D\frac{\partial}{\partial x} \left( \Delta U'(z(s),s) e^{-\beta U_{\lambda^*(s)}(z(s))} \right)}{e^{-\beta U_{\lambda^*(s)}(z(s))}}\right ) \nonumber \\
& = &  \int_0^t ds \left( \frac{\partial U_{\lambda^*(s)}(z(s)) }{\partial s} 
	+ \beta^2 D \Delta U'(z(s),s) U'_{\lambda^*(s)}(z(s)) - \beta D \Delta U''(z(s),s) \right),
\label{eq:Brownian_mod_work}
\end{eqnarray}
Using this expression for $W^*_t$ in Eq. \ref{eq:JarzPP_FK}, we can now estimate the free energy difference $F_{\lambda^*(t)}-F_{\lambda(0)}$ from trajectories generated in the process in which external parameter is switched according to the protocol $\Lambda=\lambda(t)$.

For $N$ dimensions indexed by $\alpha$, the Smoluchowski equation is,
\begin{eqnarray}
\frac{\partial f}{\partial t} = \sum_{\alpha}^N \frac{\partial}{\partial x_\alpha} \left( \frac{1}{\zeta_\alpha} \frac{\partial U_{\lambda(t)}(\{x_{\alpha'}\})}{\partial x_\alpha} f \right) + \sum_{\alpha}^N D_\alpha \frac{\partial^2}{\partial x_\alpha^2}f,
\end{eqnarray}
where $D_\alpha^{-1} = \beta \zeta_\alpha$ is the diffusion coefficient in dimension $\alpha$.  Following steps analogous to those above, we obtain the modified work,
\begin{eqnarray}
\mathcal W^*_t 
& = &  \int_0^t ds \left( \frac{\partial U_{\lambda^*(s)}}{\partial s} 
	+ \beta^2 \sum_{\alpha}^N D_\alpha \frac{\partial \Delta U}{\partial x_\alpha} \frac{\partial U_{\lambda^*(s)}}{\partial x_\alpha} - \beta \sum_{\alpha}^N D_\alpha \frac{\partial^2 \Delta U}{\partial x_\alpha^2} \right),
\label{eq:Brownian_mod_work_2D}
\end{eqnarray}
where all $U$ are implicitly functions of the position $\{x_{\alpha'}(s)\}$ at time $s$, and $\Delta  U$ is also a function of $s$.

\section{\label{sec:ISamp}Importance Sampling Formalism}

Section \ref{sec:PP} is not the first description of protocol postprocessing; 
it was preceded by a formalism based on importance sampling. \cite{Minh2009}
In this section, we describe the previous formalism in the current notation and compare it with the present results.

Explicitly in terms of path integrals, we may rewrite Eq. \ref{eq:NEW} as,
\begin{eqnarray}
e^{-\beta F_{\Lambda^*_t} } = \left< e^{-\beta W^*_t} \right>_{\Lambda^*}
\equiv 
\frac{ \int dZ ~ e^{-\beta W^*_t} \rho_{\Lambda^*}[Z] }{ \int dZ ~ \rho_{\Lambda^*}[Z] }
\end{eqnarray}
where $W^*_t \equiv  \int_0^t ds \left( \frac{ \partial H_{\lambda^*(s)}(z(s)) }{\partial s} \right)$ denotes the work performed on the system as it evolves along a particular trajectory in which the external parameter is changed according to the protocol $\Lambda^*$, $\rho_{\Lambda^*}[Z]$ is the probability density associated with the trajectory $Z$, and $dZ$ is a metric over paths. 

Now suppose that the external parameter is changed according to the protocol $\lambda(t)$ for which the associated probability  density of a trajectory $Z$ is $\rho_{\Lambda}(Z)$.
The same free energy difference may be computed by estimating different path integrals, \cite{Ytreberg2004, Minh2009}
\begin{eqnarray}
e^{-\beta F_{\Lambda^*_t} } = 
\frac{ \int dZ ~ e^{-\beta W^*_t} \left( \frac{\rho_{\Lambda^*}[Z]}{\rho_\Lambda[Z]} \right) \rho_\Lambda[Z] }{ \int dZ ~ \left( \frac{\rho_{\Lambda^*}[Z]}{\rho_\Lambda[Z]} \right) \rho_\Lambda[Z] } 
\equiv
\frac{ \left< r e^{-\beta W^*_t} \right>_\Lambda }{ \left< r \right>_\Lambda }
\label{eq:JarzPP_ISamp}
\end{eqnarray}
where $r = \rho_{\Lambda^*}[Z]/\rho_{\Lambda}[Z]$ is the ratio of densities.  If the two protocols sampling are equivalent, then $r = 1$.

This expression differs from Eq. \ref{eq:JarzPP_FK} in that it includes two expectations, the definitions of work are different, and it requires a ratio of probabilities, $r$.  The ratio is different from a ``modification'' to the work term.  For example, in overdamped Langevin dynamics, this ratio is, \cite{MinhAdib2009, Minh2009}
\begin{eqnarray}
r = \exp \left[ - \frac{\beta}{2} \left( \Delta U(z(t),t) 
	+ \int_0^t ds ~ \left( \frac{\beta D \Delta U'(z(s),s)^2}{2} - D \Delta U''(z(s),s) - \frac{ \partial \Delta U(z(s),s)}{\partial s} \right) \right) \right]
\end{eqnarray}
Now suppose that we break down $\mathcal W^*_t$ in Eq. \ref{eq:Brownian_mod_work}, into one term with $W^*_t$ and a ``modification'' term.  If we multiply this modification term by $- \beta$ and take the exponent, we obtain a term which is used similarly to $r$,
\begin{eqnarray}
\exp \left[ - \beta \left( \int_0^t ds \left( \beta D \Delta U'(z(s),s)U'_{\lambda^*(s)}(z(s))
- D \Delta U''(z(s),s) \right) \right) \right],
\end{eqnarray}
but is quite distinct.  

For multiple dimensions of overdamped Langevin dynamics, the ratio is,
\begin{eqnarray}
r = \exp \left[ - \frac{\beta}{2} \left( \Delta U(\{x_\alpha(t)\},t) 
	+ \int_0^t ds ~ \left( \sum_\alpha^N \left( \frac{\beta D_\alpha}{2} \left( \frac{\partial \Delta U}{\partial x_\alpha} \right)^2 - D_\alpha \frac{\partial^2 \Delta U}{\partial x_\alpha^2} \right) - \frac{ \partial \Delta U}{\partial s} \right) \right) \right]
\end{eqnarray}
where again, $\Delta U$ are implicitly functions of $\{x_{\alpha'}(s)\}$ and time $s$.
As in the 1D case, this expression is not equivalent to the Feynman-Kac estimator.

In later sections, we will describe several advantages of the new formalism.

\section{\label{sec:lag}Dissipation and Lag}
As protocol postprocessing is merely another mathematical formalism for computing free energies, 
there is no \emph{a priori} reason to expect that it will perform any better or worse than 
the usual nonequilibrium work estimator, Eq. \ref{eq:NEW}.
For clever choices of the analysis protocol, however, we can show that Eq. \ref{eq:JarzPP_FK} leads to a highly efficient estimator for $F_{\Lambda^*_t}$.

\subsection{\label{sec:analytical}Exactly solved models}

Suppose that we construct a ``perfect'' analysis protocol $\lambda^*(t)$ whose instantaneous equilibrium density is equivalent to the nonequilibrium density, 
so that $f(z,t)= f^{eq}_{\lambda^*(t)}(z)$, where  
$ f^{eq}_{\lambda}(z)\equiv Q_{\lambda}^{-1} e^{-\beta H_{\lambda}(z)}= e^{-\beta (H_{\lambda}(z)-F_{\lambda})}$ denotes the equilibrium distribution corresponding to $\beta^{-1}$ and $\lambda$. 
When a perfect analysis protocol is used, then
\begin{equation}
\label{eq:perfectwork}
\mathcal W_t^* =F_{\Lambda^*_t}
\end{equation}
for \emph{every} trajectory!  This may be seen by first substituting 
$f(z,t)=e^{-\beta (H_{\lambda^*(t)}(z)-F_{\lambda^*_t}) }$ in the evolution equation,
\begin{equation}
\frac{\partial f(z,t)}{\partial t} 
= \mathcal L_\lambda(t) \cdot f(z,t)
= \mathcal L_{\lambda^*(t)} \cdot f(z,t) + \mathcal A(t) \cdot f(z,t)
\end{equation}
where we have used Eq. \ref{eq:separation}.
Since $\mathcal L_{\lambda^*(t)} \cdot f(z,t) = 0$ for this $f(z,t)$, we obtain,
\begin{eqnarray}
 -\beta \left( \frac{\partial H_{\lambda^*(t)}(z)}{\partial t}-\frac{\partial F_{\lambda^*_t}}{\partial t} \right) 
 e^{-\beta H_{\lambda^*(t)}(z)-F_{\lambda^*_t}}
& = & \mathcal A(t) \cdot e^{-\beta (H_{\lambda^*(t)}(z)-F_{\lambda^*_t})} \nonumber \\
\frac{\partial F_{\lambda^*_t}}{\partial t}
& = & \frac{\partial H_{\lambda^*(t)}(z)}{\partial t}+\frac {\mathcal A(t) \cdot e^{-\beta (H_{\lambda^*(t)}(z))}}{ \beta e^{-\beta H_{\lambda^*(t)}(z)} } 
\end{eqnarray}
By substituting this into the modified work, Eq. \ref{eq:mod_work} and integrating, we obtain Eq. \ref{eq:perfectwork}. As this equation is valid for every trajectory, Eq. \ref{eq:JarzPP_FK} is a \emph{zero variance} estimator of $F_{\Lambda^*_t}$. 
As a demonstration of this principle, consider two exactly solved \cite{Minh2009} models:
a Brownian particle in a harmonic oscillator that either (i) has its center moving at a constant velocity, or (ii) has a time-dependent natural frequency.  
In both cases, the potential has the general time-dependent form $U_{\lambda(t)}(z) = \frac{k(t)}{2}(z-\bar{z}(t))^2$ where the vector $\lambda(t)= \{ k(t), \bar{z}(t) \} $ denotes the set of external parameters. The  Smoluchowski equation describing the evolution of the phase space density $f(z,t)$ can be solved to give \cite{MinhAdib2009, Minh2009}
\begin{eqnarray}
f(z,t) = \sqrt{ \frac{\beta k_T(t) }{ 2 \pi } } e^{\frac{ -\beta k_T(t)}{2}(z-z_T(t))^2 },
\end{eqnarray}
where $z_T=\langle z \rangle $, $k_T(t)=1/(\langle z^2-{\langle z \rangle}^2)$, and $\langle \dots \rangle$ denotes an average over the distribution $f(z,t)$.   
In case (i), the spring coefficient $k(t)$ is  held fixed at $k$ while $\bar{z}(t)$ is switched according to $\bar{z}(t) = vt$ ($\lambda(t)=\{k,vt\}$). In this case, the free energy difference is always zero and $k_T(t)$ is a constant, $k$.  The most typical path is,
\begin{equation}
z_T(t)  = vt - \frac{v}{\beta D k} (1 - e^{-\beta D kt}).
\label{eq:movingHO_x_T}
\end{equation}
In case (ii), $\bar{z}(t)$ is held fixed at $\bar{z}(t)=0$ and the spring coefficient $k(t)$ is switched according to $k(t) =vt$ ($\lambda(t)=\{vt,0\}$).  In this case, $z_T(t) = 0$, and 
\begin{equation}
k_T(t) = \frac{ k(0) e^{2\beta D \int_0^t ds~k(s)}}{1+2 \beta Dk(0)  \left[\int_0^t du~e^{2 \beta D \int_0^u ds~k(s)} \right]}.
\label{eq:changingHO_k_T}
\end{equation}

In either case, we may choose the analysis protocol $\lambda^*(t)\equiv \{k_T(t),z_T(t)\}$ such that $U_{\lambda^*(t)}(z) = \frac{k_T(t)}{2}(z-{z_T(t)})^2 $. 
With this choice, the Boltzmann distribution corresponding to the analysis protocol is equal to $f(z,t)$. 
Hence, the modified work calculated from Eq. \ref{eq:Brownian_mod_work} is always equal to the free energy difference $F_{\Lambda^*_t}$.
In contrast, the importance sampling form of protocol postprocessing yields different work values for each trajectory.

\subsection{Dissipation Bounds Lag}

In general, it is not feasible to find a perfect analysis protocol.  Indeed, in most cases, the nonequilibrium densities $f(z,t)$ will not belong to the family of equilibrium distributions indexed by $\lambda$, $f^{eq}_{\lambda}$.
However, Eq. \ref{eq:perfectwork} suggests that efficient estimators of free energy energies can be obtained if we can find an analysis protocol $\lambda^*(t)$ such that $ f^{eq}_{\lambda^*(t)}(z)$ closely resembles the nonequilibrium density $f(z,t)$. In the following paragraphs, we will make this argument more rigorous.

The convergence of the protocol postprocessing form of Jarzynski's equality will depend on a criterion analogous to that in the original form: obtaining trajectories in which the modified work, $\mathcal W^*_t$, is less than the free energy difference, $F_{\Lambda^*_t}$. \cite{Jarzynski2006, Kofke2006}
Chances of obtaining such trajectories are improved when the average dissipation, 
$\mathcal W^*_{d} \equiv \left< \mathcal W^*_t \right>_\Lambda - F_{\Lambda^*_t}$, is small. \cite{Jarzynski2006, Kofke2006}
This dissipation can be related to an information theoretic measure of overlap between the distributions $f(z,t)$ describing the state of the system and the equilibrium state corresponding to the $\lambda^*(t)$, $ f^{eq}_{\lambda^*(t)}(z)$. To obtain this relation, we note that the properties of the delta function enable the path-ensemble average in Eq. \ref{eq:weighted_density2} to be written as,
\begin{eqnarray}
\left< \delta(z - z(t)) e^{-\beta \mathcal W^*_t} \right>_{\Lambda} = \left< \delta (z - z(t)) \right>_\Lambda \left< e^{-\beta \mathcal W^*_t}\right>_{\Lambda; (z,t)},
\end{eqnarray}
where the double subscript $\left< ... \right>_{\Lambda; (z,t)}$ indicates a path-ensemble average for trajectories driven with the protocol $\Lambda$ and which pass through $z$ at time $t$.  Since the nonequilibrium density at time $t$ is $f(z,t) = \left< \delta (z - z(t)) \right>_\Lambda$, we may rearrange Eq. \ref{eq:weighted_density2} to obtain,
\begin{eqnarray}
\frac{f(z,t)}{f^{eq}_{\lambda^*(t)}(z)} = \frac{ e^{-\beta F_{\Lambda^*_t}} }{ \left< e^{-\beta \mathcal W^*_t}\right>_{\Lambda; (z,t)} }.
\end{eqnarray}

As in Ref. \cite{Vaikuntanathan2009}, we then take the logarithm of both sides of the equation, invoke Jensen's inequality, multiply both sides by $f_{\lambda^*(t)}(z)$, and integrate over $z$.  Our final result is,
\begin{eqnarray}
\left< \mathcal W^*_t \right>_\Lambda - F_{\Lambda^*_t} & \geq & \beta^{-1} \int dz ~ f(z,t) \ln \frac{f_{}(z,t)}{f^{eq}_{\lambda^*(t)}(z)} \nonumber \\
& \equiv & \beta^{-1} D[f_{}(z,t)||f^{eq}_{\lambda^*(t)}(z)],
\label{eq:diss_and_lag}
\end{eqnarray}
where $D[f_{}(z,t)||f^{eq}_{\lambda^*(t)}(z)]$ is the Kullback-Leibler divergence, or relative entropy, between the nonequilibrium density and the equilibrium density corresponding to the analysis protocol.  The relative entropy is zero when two distributions are identical and grows larger when they diverge \cite{Cover1991}.
Eq. \ref{eq:diss_and_lag} suggests, but does not prove (the inequality goes the wrong way), that a reasonable strategy for reducing dissipation and improving the convergence of the free energy estimator is to choose an analysis protocol in which the ``analysis" density closely resembles the evolving state of the system. 

\section{General Case}

Based on the results in Section \ref{sec:lag}, we speculate that a reasonable strategy for minimizing dissipation and improving the efficiency of the free energy estimator is to choose an analysis protocol $\Lambda^* \equiv \lambda^*(t)$ so that the Kullback-Leibler divergence $D[f_{}(z,t)||f^{eq}_{\lambda^*(t)}(z)]$ is small for all $t$.
Obtaining such a protocol will usually entail a search over the space of $\lambda$ to find an equilibrium distribution $f_{\lambda}^{eq}(z)$ which is similar to $f(z,t)$.
While the nonequilibrium distribution is not analytically tractable for most systems, it is possible to use sampled trajectories to compare the relative entropy between $f(z,t)$ and $f_{\lambda}^{eq}(z)$ for different values of $\lambda$. 
Specifically, given a set of trajectories $\mathcal \{Z_1, Z_2, ..., Z_{N_s} \}$ and several candidate values of $\lambda$, the relative entropy $D[f(z,t)||f^{eq}_{\lambda}(z)]$ is minimized by the parameter vector $\lambda$ that minimizes $\langle H_{\lambda}(z) \rangle_{f(z,t)}-  F_{\lambda}$, which may be estimated by the sample average, \cite{Minh2009}
\begin{eqnarray}
D_{Test}(\mathcal Z, t) = \frac{1}{N_s} \left[ \sum_{n=1}^{N_s} H_{\lambda}(z_n(t)) \right] -  F_{\lambda}.
\label{eq:DKLestimator}
\end{eqnarray}
where $z_n(t)$ denotes the state of system in phase space at time $t$ as it evolves along the trajectory $Z_n$.
(Note that in $D[f(z,t)||f^{eq}_{\lambda}(z)]$, the integral $\int dz~f(z,t)\ln f^{}_{}(z,t)$  does not depend on $\lambda$.)
A reasonable choice for the search space of $\lambda$ is the range of the sampling protocol $\Lambda$.
This choice has the advantage that $F_\lambda - F_{\lambda(0)}$ may be estimated via Jarzynski's equality; for distributions that are not accessed during the sampling protocol, it may be more difficult to estimate corresponding free energies.

As noted in Section \ref{sec:FK}, the flexibility in choosing $\Lambda^*$ means that the free energy $F_{\Lambda^*_t}$ may be different from $F_{\Lambda_t}$.  Indeed, unless there is no lag, an analysis protocol which minimizes the lag will \emph{always} have different states than the sampling protocol.  
Since we are typically interested in free energies between the end states of the sampling protocol ($A\equiv \lambda(0)$ and $B\equiv \lambda(T)$), this discrepancy was addressed by introducing an adaptive algorithm, nonequilibrium density-dependent sampling (NEDDS). \cite{Minh2009}
NEDDS is equally applicable to the current formalism.

In brief, NEDDS entails running all $N_s$ desired simulations of the nonequilibrium process simultaneously.
The sampling protocol initially involves an interpolation between the desired end states $A$ and $B$.
After reaching state $B$, the protocol extrapolates past it until an adaptively determined stopping time.
(While such an extrapolation may not always be physically meaningful, it is nearly always computationally feasible.)
Without loss of generality, let us assume that $A < B$.
The stopping time is decided by performing the following calculations while the simulations are in progress:
\begin{enumerate}
\item The free energy difference, $F_{\lambda(t)}-F_{\lambda(0)}$, between the initial and instantaneous state at the current time step, $t$, is estimated using Eq. \ref{eq:NEW}.
\item $D_{Test}$ is evaluated with $\lambda$ values from the current state and all preceding states using Eq. \ref{eq:DKLestimator}.
\item If the choice of $\lambda$ that minimizes $D_{Test}$, $\lambda^{min}$, is between $A$ and $B$, $A<\lambda^{min}<B$, then it is appended to the analysis protocol, $\lambda^*(t) = \lambda^{min}$.
Otherwise, if it is at or beyond $B$, $\lambda^{min} \geq B$, then the final value of the analysis protocol is set to $B$, $\lambda^*(t)=B$.
\item Lastly, $W_t^*$ is incremented and $F_{\Lambda^*_t}$ is evaluated by protocol postprocessing.
\end{enumerate}
This procedure ensures that protocol postprocessing estimators can compute the free energy difference between the states $A$ and $B$.

\section{Model Systems}

We now demonstrate NEDDS with protocol postprocessing (both importance sampling and Feynman-Kac) formalisms and compare its efficiency to standard sample mean estimates from Jarzynski's equality, Eq. \ref{eq:NEW}, on a few toy model systems.
First, consider an overdamped Brownian particle evolving on an one-dimensional surface, $U(z,\lambda) = z^4 - 16\lambda z^2$, as studied by Sun. \cite{Sun2003}
In this system, the free energy difference between the states $\lambda = 0$ and $\lambda = 1$ at $\beta=1$ was analytically found to be $F_{\lambda = 1} - F_{\lambda = 0} = -62.9407$. \cite{Oberhofer2005}

As described, \cite{Minh2009} simulations of nonequilibrium driven processes were performed in which $\lambda$ was switched between 0 and 1 according to the equation of motion,
\begin{equation}
z_{j+1} = z_{j} - D \Delta t U_j' + \sqrt{2D\Delta t} R_j,
\label{eq:BD}
\end{equation}
where $z_j$ is the position at time $j \Delta t$, $D = 1$ is the diffusion coefficient, $\Delta t = 0.0001$ is the time step, and $R_j$ is a standard normal random variable.
$\lambda$ was incremented at each time step by $v \Delta t$.  NEDDS was used to obtain the analysis protocol $\lambda^*(t)$ concluding at $\lambda^*(t) = 1$, and the free energy difference $F_{\lambda = 1} - F_{\lambda = 0}$ was computed using either Eq. \ref{eq:JarzPP_FK} or Eq. \ref{eq:JarzPP_ISamp}. 
For comparison, the standard Jarzynski estimate was applied to two types of simulations taking the same amount of simulation time as the analysis protocol obtained from NEDDS: either (i) $\lambda$ was switched between 0 and 1 at a slower velocity, or (ii) the NEDDS analysis protocol was used as a new sampling protocol.

\begin{figure}
\begin{center}
\includegraphics{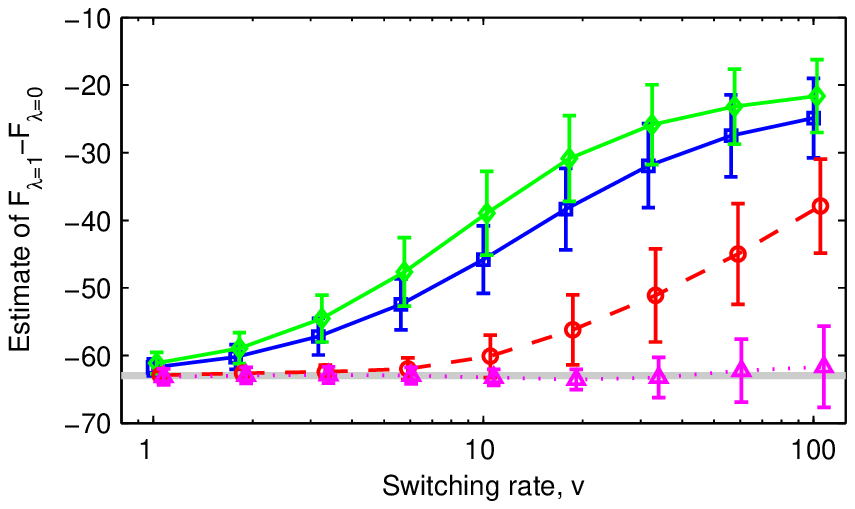}
\caption{\label{fig:Sun_FE}
Comparison of free energy estimates for Sun's system:
NEDDS simulations were analyzed with importance sampling, Eq. \ref{eq:JarzPP_ISamp} (circles), or the Feynman-Kac formalism, Eq. \ref{eq:JarzPP_FK} (triangles).
Standard Jarzynski estimates, Eq. \ref{eq:NEW} (squares), were performed on slower simulations with the same total time as the NEDDS simulations or by using the analysis protocol as a new sampling protocol (diamonds).
The symbols indicate the mean and error bars indicate the standard deviation of 10000 estimates, each based on 50 trajectories.
The simulation time step was $\Delta t = 0.001$ and the rate $v$ indicates that $\lambda$ was incremented by $v \Delta t$ at each time step of the NEDDS simulations.
While the switching rates are equivalent, some plots are slightly offset to prevent error bar overlap.
The exact free energy is shown as a shaded line.
}
\end{center}
\end{figure}

While the importance sampling formalism was found to be an improvement over the standard form of Jarzynski's equality, \cite{Minh2009} 
we find that the estimator based on Eq. \ref{eq:JarzPP_FK} is even better (Fig. \ref{fig:Sun_FE}).  Even for the fastest switching rates, where dissipation is expected to be high, the systematic bias is largely eliminated.
No benefit was found from using the analysis protocol from NEDDS as a new sampling protocol; in fact, the bias was worse than with the constant velocity protocol.

We also performed similar tests on another one-dimensional surface, 
$U(z,\lambda) = (5z^3 - 10z + 3)z + \frac{15}{2} (z - \lambda(t))^2$, 
first described by Hummer. \cite{Hummer2007}
Hummer's surface, a double well potential that includes a harmonic bias, 
mimics the setup of a single-molecule pulling experiment, 
and hence has been used to demonstrate estimators of free energies \cite{MinhAdib2008, MinhChodera2009} and other quantities \cite{MinhChodera2010} in the context of these experiments.  The simulations were performed using the same equation of motion, diffusion coefficient, and time step as described above for Sun's system.  $\lambda$ was switched between -1.5 and 1.5.

\begin{figure}
\begin{center}
\includegraphics{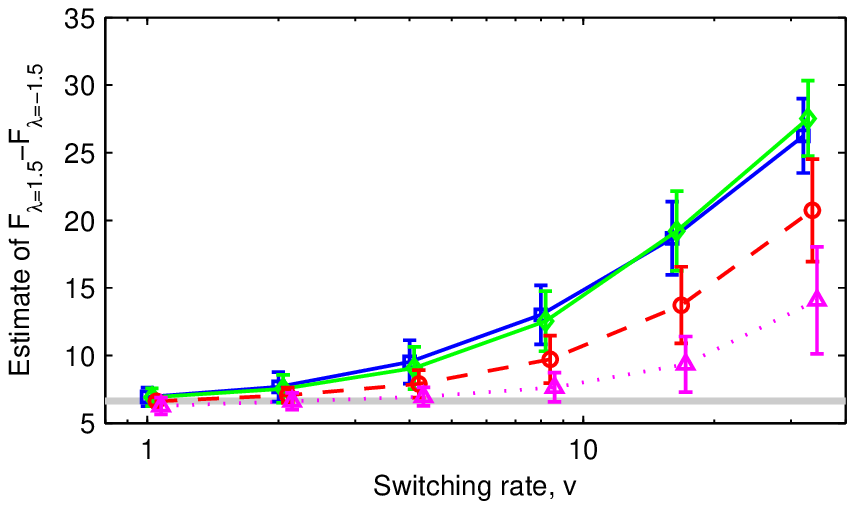}
\caption{\label{fig:Hummer_FE}
Comparison of free energy estimates for Hummer's system.  The caption for Fig. \ref{fig:Sun_FE} applies here, except that the potential is Hummer's rather than Sun's and each estimate is based on 250 trajectories.
}
\end{center}
\end{figure}

The performance trends with Hummer's system are similar to those with Sun's (Fig. \ref{fig:Hummer_FE}).  Results from the standard form of Jarzynski's equality are more biased than with NEDDS and the importance sampling formalism, which in turn is more biased than the Feynman-Kac formalism.  In contrast to Sun's system, however, the estimates from Eq. (\ref{eq:JarzPP_FK}) are noticeably biased at the fastest switching rates.  Another distinction between the trends from the two systems is that results from using a constant velocity protocol and using the analysis protocol as a new sampling protocol are rather similar.

\begin{figure}
\begin{center}
\includegraphics{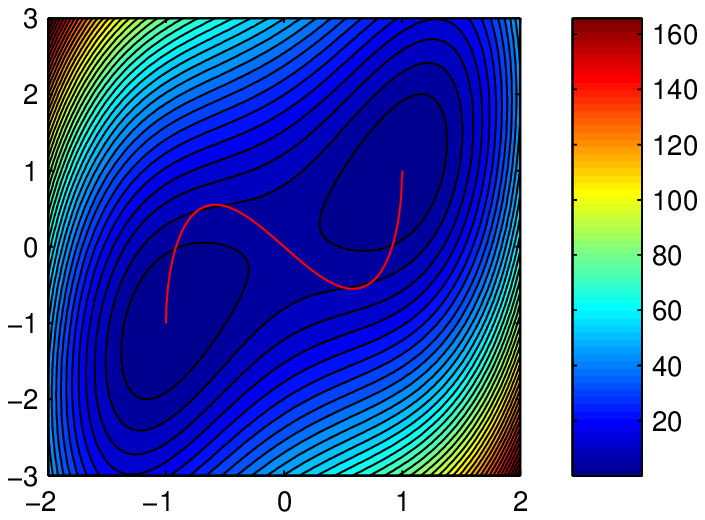}
\caption{\label{fig:potential_2D}
Potential energy surface for a 2D system.  The contour plot is of $5(x^2 - 1)^2 + 5(x-y)^2$ and the red line traces the equilibrium position of the harmonic bias $\frac{15}{2} (x+\cos(\pi \lambda))^2 + \frac{15}{2} (y+1-\sin(2 \pi \lambda)-2\lambda)^2$ as $\lambda$ goes from 0 (left) to 1 (right).
}
\end{center}
\end{figure}

As a final demonstration, we consider a two-dimensional surface,
\begin{equation}
U(x,y,\lambda) = 5(x^2 - 1)^2 + 5(x-y)^2 
+ \frac{15}{2} (x+\cos(\pi \lambda))^2 + \frac{15}{2} (y+1-\sin(2 \pi \lambda)-2\lambda)^2,
\label{eq:2D_surface}
\end{equation}
in which $\lambda$ dictates the progress of a harmonic bias along a curve (Fig. \ref{fig:potential_2D}).
Simulations were performed as with the 1D potentials, using Eq. \ref{eq:BD} along each dimension, as well as the same diffusion coefficient and time step.  $\lambda$ was switched between 0 and 1.

\begin{figure}
\begin{center}
\includegraphics{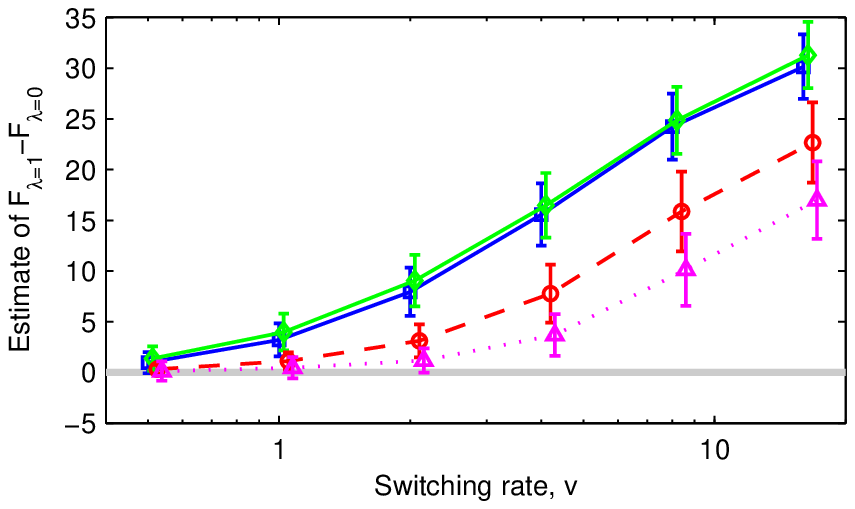}
\caption{\label{fig:2D_FE}
Comparison of free energy estimates for a 2D system.
The caption for Fig. \ref{fig:Sun_FE} applies here, except that the potential is Eq. \ref{eq:2D_surface} rather than Sun's system, each estimate is based on 250 trajectories, and multidimensional versions of the importance sampling and Feynman-Kac formalisms were used.
}
\end{center}
\end{figure}

The performance trends in this system are the same as in Hummer's system (Fig. \ref{fig:2D_FE}).

\section{Discussion and Conclusion}

We have presented a method for analyzing nonequilibrium trajectories which borrows from a similar philosophy as previous work \cite{Minh2009} but is based on a distinct mathematical formalism.  The new formalism has the advantages that it analytically is a zero-variance estimator if a ``perfect'' analysis protocol is obtained, and it improves the convergence of free energy estimates in all our tested model systems.  Further tests on more complex multidimensional systems are a potential future research direction.  

We expect that protocol postprocessing will be most useful when 
(i) there is little phase space overlap between the end states of interest 
(otherwise free energy differences can be computed without nonequilibrium work identities),
(ii) estimates of $\Delta F$ from the nonequilibrium work relation suffer from poor convergence 
for a given nonequilbrium process in which the system is driven between the end states of interest and
(iii) it is reasonable to speculate that the nonequilibrium driven process has a nonequilibrium
density $f(z,t)$ that always resembles an equilibrium density 
$f^{eq}_{\lambda}$ parameterized by a $\lambda$ vector along the
protocol.  
Exact convergence properties, of course, will depend on the system.

The observed convergence benefits hint that many improved sampling and analysis algorithms based on nonequilibrium driven processes still remain to be discovered.

\section{Acknowledgments}

We thank Andy Ballard, John Chodera, and Christopher Jarzynski for helpful comments on the manuscript.  D. Minh is funded by a Director's Postdoctoral Fellowship at Argonne and S. Vaikuntanathan acknowledges support from the National Science Foundation (USA) under CHE-0841557 and the University of Maryland, College Park.


\begin{thebibliography}{10}%
\makeatletter
\providecommand \@ifxundefined [1]{%
 \ifx #1\undefined \expandafter \@firstoftwo
 \else \expandafter \@secondoftwo
\fi
}%
\providecommand \@ifnum [1]{%
 \ifnum #1\expandafter \@firstoftwo
 \else \expandafter \@secondoftwo
\fi
}%
\providecommand \enquote [1]{``#1''}%
\providecommand \bibnamefont  [1]{#1}%
\providecommand \bibfnamefont [1]{#1}%
\providecommand \citenamefont [1]{#1}%
\providecommand\href[0]{\@sanitize\@href}%
\providecommand\@href[1]{\endgroup\@@startlink{#1}\endgroup\@@href}%
\providecommand\@@href[1]{#1\@@endlink}%
\providecommand \@sanitize [0]{\begingroup\catcode`\&12\catcode`\#12\relax}%
\@ifxundefined \pdfoutput {\@firstoftwo}{%
 \@ifnum{\z@=\pdfoutput}{\@firstoftwo}{\@secondoftwo}%
}{%
 \providecommand\@@startlink[1]{\leavevmode}%
 \providecommand\@@endlink[0]{}%
}{%
 \providecommand\@@startlink[1]{%
  \leavevmode
  \pdfstartlink
   attr{/Border[0 0 1 ]/H/I/C[0 1 1]}%
   user{/Subtype/Link/A<</Type/Action/S/URI/URI(#1)>>}%
  \relax
 }%
 \providecommand\@@endlink[0]{\pdfendlink}%
}%
\providecommand \url  [0]{\begingroup\@sanitize \@url }%
\providecommand \@url [1]{\endgroup\@href {#1}{\urlprefix}}%
\providecommand \urlprefix [0]{URL }%
\providecommand \Eprint[0]{\href }%
\@ifxundefined \urlstyle {%
  \providecommand \doi [1]{doi:\discretionary{}{}{}#1}%
}{%
  \providecommand \doi [0]{doi:\discretionary{}{}{}\begingroup
  \urlstyle{rm}\Url }%
}%
\providecommand \doibase [0]{http://dx.doi.org/}%
\providecommand \Doi[1]{\href{\doibase#1}}%
\providecommand \selectlanguage [0]{\@gobble}%
\providecommand \bibinfo [0]{\@secondoftwo}%
\providecommand \bibfield [0]{\@secondoftwo}%
\providecommand \translation [1]{[#1]}%
\providecommand \BibitemOpen[0]{}%
\providecommand \bibitemStop [0]{}%
\providecommand \bibitemNoStop [0]{.\EOS\space}%
\providecommand \EOS [0]{\spacefactor3000\relax}%
\providecommand \BibitemShut [1]{\csname bibitem#1\endcsname}%
\bibitem{Chipot2007}%
  \BibitemOpen
  \bibfield{author}{%
  \bibinfo {author} {\bibfnamefont{C.}~\bibnamefont{Chipot}}\ and\ \bibinfo
  {author} {\bibfnamefont{A.}~\bibnamefont{Pohorille}},\ }%
  \emph{\bibinfo {title} {Free Energy Calculations}}\ (\bibinfo {publisher}
  {Springer, Berlin},\ \bibinfo {year} {2007})\BibitemShut{NoStop}%
\bibitem{Jarzynski1997a}%
  \BibitemOpen
  \bibfield{author}{%
  \bibinfo {author} {\bibfnamefont{C.}~\bibnamefont{Jarzynski}},\ }%
  \bibfield{journal}{%
  \bibinfo {journal} {Phys. Rev. Lett.}\ }%
  \textbf{\bibinfo {volume} {78}},\ \bibinfo {pages} {2690} (\bibinfo {year}
  {1997})\BibitemShut{NoStop}%
\bibitem{Jarzynski1997b}%
  \BibitemOpen
  \bibfield{author}{%
  \bibinfo {author} {\bibfnamefont{C.}~\bibnamefont{Jarzynski}},\ }%
  \bibfield{journal}{%
  \bibinfo {journal} {Phys. Rev. E}\ }%
  \textbf{\bibinfo {volume} {56}},\ \bibinfo {pages} {5018} (\bibinfo {year}
  {1997})\BibitemShut{NoStop}%
\bibitem{Crooks1998}%
  \BibitemOpen
  \bibfield{author}{%
  \bibinfo {author} {\bibfnamefont{G.~E.}\ \bibnamefont{Crooks}},\ }%
  \bibfield{journal}{%
  \bibinfo {journal} {J. Stat. Phys.}\ }%
  \textbf{\bibinfo {volume} {90}},\ \bibinfo {pages} {1481} (\bibinfo {year}
  {1998})\BibitemShut{NoStop}%
\bibitem{Crooks1999}%
  \BibitemOpen
  \bibfield{author}{%
  \bibinfo {author} {\bibfnamefont{G.~E.}\ \bibnamefont{Crooks}},\ }%
  \bibfield{journal}{%
  \Doi{10.1103/PhysRevE.60.2721}{\bibinfo {journal} {Phys. Rev. E}}\ }%
  \textbf{\bibinfo {volume} {60}},\ \bibinfo {pages} {2721} (\bibinfo {month}
  {Sep}\ \bibinfo {year} {1999})\BibitemShut{NoStop}%
\bibitem{Crooks2000}%
  \BibitemOpen
  \bibfield{author}{%
  \bibinfo {author} {\bibfnamefont{G.~E.}\ \bibnamefont{Crooks}},\ }%
  \bibfield{journal}{%
  \bibinfo {journal} {Phys. Rev. E}\ }%
  \textbf{\bibinfo {volume} {61}},\ \bibinfo {pages} {2361} (\bibinfo {year}
  {2000})\BibitemShut{NoStop}%
\bibitem{Frenkel2002}%
  \BibitemOpen
  \bibfield{author}{%
  \bibinfo {author} {\bibfnamefont{D.}~\bibnamefont{Frenkel}}\ and\ \bibinfo
  {author} {\bibfnamefont{B.}~\bibnamefont{Smit}},\ }%
  \emph{\bibinfo {title} {Understanding Molecular Simulation}}\ (\bibinfo
  {publisher} {Elsevier},\ \bibinfo {year} {2002})\ pp.\ \bibinfo {pages}
  {183--189}\BibitemShut{NoStop}%
\bibitem{Gore2003}%
  \BibitemOpen
  \bibfield{author}{%
  \bibinfo {author} {\bibfnamefont{J.}~\bibnamefont{Gore}}, \bibinfo {author}
  {\bibfnamefont{F.}~\bibnamefont{Ritort}},\ and\ \bibinfo {author}
  {\bibfnamefont{C.}~\bibnamefont{Bustamante}},\ }%
  \bibfield{journal}{%
  \bibinfo {journal} {Proc. Natl. Acad. Sci. U.S.A.}\ }%
  \textbf{\bibinfo {volume} {100}},\ \bibinfo {pages} {12564} (\bibinfo {year}
  {2003})\BibitemShut{NoStop}%
\bibitem{Jarzynski2006}%
  \BibitemOpen
  \bibfield{author}{%
  \bibinfo {author} {\bibfnamefont{C.}~\bibnamefont{Jarzynski}},\ }%
  \bibfield{journal}{%
  \bibinfo {journal} {Phys. Rev. E}\ }%
  \textbf{\bibinfo {volume} {73}},\ \bibinfo {pages} {046105} (\bibinfo {year}
  {2006})\BibitemShut{NoStop}%
\bibitem{Kofke2006}%
  \BibitemOpen
  \bibfield{author}{%
  \bibinfo {author} {\bibfnamefont{D.~A.}\ \bibnamefont{Kofke}},\ }%
  \bibfield{journal}{%
  \bibinfo {journal} {Mol. Phys.}\ }%
  \textbf{\bibinfo {volume} {104}},\ \bibinfo {pages} {3701} (\bibinfo {year}
  {2006})\BibitemShut{NoStop}%
\bibitem{Vaikuntanathan2008}%
  \BibitemOpen
  \bibfield{author}{%
  \bibinfo {author} {\bibfnamefont{S.}~\bibnamefont{Vaikuntanathan}}\ and\
  \bibinfo {author} {\bibfnamefont{C.}~\bibnamefont{Jarzynski}},\ }%
  \bibfield{journal}{%
  \bibinfo {journal} {Phys. Rev. Lett.}\ }%
  \textbf{\bibinfo {volume} {100}},\ \bibinfo {pages} {190601} (\bibinfo {year}
  {2008})\BibitemShut{NoStop}%
\bibitem{Vaikuntanathan2009}%
  \BibitemOpen
  \bibfield{author}{%
  \bibinfo {author} {\bibfnamefont{S.}~\bibnamefont{Vaikuntanathan}}\ and\
  \bibinfo {author} {\bibfnamefont{C.}~\bibnamefont{Jarzynski}},\ }%
  \bibfield{journal}{%
  \bibinfo {journal} {Europhys. Lett.}\ }%
  \textbf{\bibinfo {volume} {87}},\ \bibinfo {pages} {60005} (\bibinfo {year}
  {2009})\BibitemShut{NoStop}%
\bibitem{LandauLifshitz}%
  \BibitemOpen
  \bibfield{author}{%
  \bibinfo {author} {\bibfnamefont{L.}~\bibnamefont{Landau}}\ and\ \bibinfo
  {author} {\bibfnamefont{E.}~\bibnamefont{Lifshitz}},\ }%
  \emph{\bibinfo {title} {Statistical Physics}},\ \bibinfo {edition} {3rd}\
  ed.\ (\bibinfo {publisher} {Pergamon Press},\ \bibinfo {address} {Oxford},\
  \bibinfo {year} {1990})\BibitemShut{NoStop}%
\bibitem{Athenes2002}%
  \BibitemOpen
  \bibfield{author}{%
  \bibinfo {author} {\bibfnamefont{M.}~\bibnamefont{Athenes}},\ }%
  \bibfield{journal}{%
  \bibinfo {journal} {Phys. Rev. E}\ }%
  \textbf{\bibinfo {volume} {66}},\ \bibinfo {pages} {046705} (\bibinfo {year}
  {2002})\BibitemShut{NoStop}%
\bibitem{Sun2003}%
  \BibitemOpen
  \bibfield{author}{%
  \bibinfo {author} {\bibfnamefont{S.}~\bibnamefont{Sun}},\ }%
  \bibfield{journal}{%
  \bibinfo {journal} {J. Chem. Phys.}\ }%
  \textbf{\bibinfo {volume} {118}},\ \bibinfo {pages} {5769} (\bibinfo {year}
  {2003})\BibitemShut{NoStop}%
\bibitem{Atilgan2004}%
  \BibitemOpen
  \bibfield{author}{%
  \bibinfo {author} {\bibfnamefont{E.}~\bibnamefont{Atilgan}}\ and\ \bibinfo
  {author} {\bibfnamefont{S.~X.}\ \bibnamefont{Sun}},\ }%
  \bibfield{journal}{%
  \Doi{10.1063/1.1813434}{\bibinfo {journal} {J. Chem. Phys.}}\ }%
  \textbf{\bibinfo {volume} {121}},\ \bibinfo {pages} {10392} (\bibinfo {year}
  {2004})\BibitemShut{NoStop}%
\bibitem{Ytreberg2004}%
  \BibitemOpen
  \bibfield{author}{%
  \bibinfo {author} {\bibfnamefont{F.~M.}\ \bibnamefont{Ytreberg}}\ and\
  \bibinfo {author} {\bibfnamefont{D.~M.}\ \bibnamefont{Zuckerman}},\ }%
  \bibfield{journal}{%
  \bibinfo {journal} {J. Chem. Phys.}\ }%
  \textbf{\bibinfo {volume} {120}},\ \bibinfo {pages} {10876} (\bibinfo {month}
  {Jan}\ \bibinfo {year} {2004})\BibitemShut{NoStop}%
\bibitem{Athenes2004}%
  \BibitemOpen
  \bibfield{author}{%
  \bibinfo {author} {\bibfnamefont{M.}~\bibnamefont{Athenes}},\ }%
  \bibfield{journal}{%
  \bibinfo {journal} {Eur. Phys. J. B}\ }%
  \textbf{\bibinfo {volume} {38}},\ \bibinfo {pages} {651} (\bibinfo {year}
  {2004})\BibitemShut{NoStop}%
\bibitem{Oberhofer2005}%
  \BibitemOpen
  \bibfield{author}{%
  \bibinfo {author} {\bibfnamefont{H.}~\bibnamefont{Oberhofer}}, \bibinfo
  {author} {\bibfnamefont{C.}~\bibnamefont{Dellago}},\ and\ \bibinfo {author}
  {\bibfnamefont{P.}~\bibnamefont{Geissler}},\ }%
  \bibfield{journal}{%
  \bibinfo {journal} {J. Phys. Chem. B}\ }%
  \textbf{\bibinfo {volume} {109}},\ \bibinfo {pages} {6902} (\bibinfo {month}
  {Jan}\ \bibinfo {year} {2005})\BibitemShut{NoStop}%
\bibitem{Oberhofer2008}%
  \BibitemOpen
  \bibfield{author}{%
  \bibinfo {author} {\bibfnamefont{H.}~\bibnamefont{Oberhofer}}\ and\ \bibinfo
  {author} {\bibfnamefont{C.}~\bibnamefont{Dellago}},\ }%
  \bibfield{journal}{%
  \bibinfo {journal} {Comput. Phys. Commun.}\ }%
  \textbf{\bibinfo {volume} {179}},\ \bibinfo {pages} {41} (\bibinfo {year}
  {2008})\BibitemShut{NoStop}%
\bibitem{Minh2009}%
  \BibitemOpen
  \bibfield{author}{%
  \bibinfo {author} {\bibfnamefont{D.~D.~L.}\ \bibnamefont{Minh}},\ }%
  \bibfield{journal}{%
  \bibinfo {journal} {J. Chem. Phys.}\ }%
  \textbf{\bibinfo {volume} {130}},\ \bibinfo {pages} {204102} (\bibinfo {year}
  {2009})\BibitemShut{NoStop}%
\bibitem{Kac1949}%
  \BibitemOpen
  \bibfield{author}{%
  \bibinfo {author} {\bibfnamefont{M.}~\bibnamefont{Kac}},\ }%
  \bibfield{journal}{%
  \bibinfo {journal} {Trans. Am. Math Soc.}\ }%
  \textbf{\bibinfo {volume} {65}},\ \bibinfo {pages} {1Ð13} (\bibinfo {year}
  {1949})\BibitemShut{NoStop}%
\bibitem{Hummer2001a}%
  \BibitemOpen
  \bibfield{author}{%
  \bibinfo {author} {\bibfnamefont{G.}~\bibnamefont{Hummer}}\ and\ \bibinfo
  {author} {\bibfnamefont{A.}~\bibnamefont{Szabo}},\ }%
  \bibfield{journal}{%
  \bibinfo {journal} {Proc. Natl. Acad. Sci. U.S.A.}\ }%
  \textbf{\bibinfo {volume} {98}},\ \bibinfo {pages} {3658} (\bibinfo {year}
  {2001})\BibitemShut{NoStop}%
\bibitem{Hummer2005}%
  \BibitemOpen
  \bibfield{author}{%
  \bibinfo {author} {\bibfnamefont{G.}~\bibnamefont{Hummer}}\ and\ \bibinfo
  {author} {\bibfnamefont{A.}~\bibnamefont{Szabo}},\ }%
  \bibfield{journal}{%
  \bibinfo {journal} {Acc. Chem. Res.}\ }%
  \textbf{\bibinfo {volume} {38}},\ \bibinfo {pages} {504} (\bibinfo {year}
  {2005})\BibitemShut{NoStop}%
\bibitem{Ge2008}%
  \BibitemOpen
  \bibfield{author}{%
  \bibinfo {author} {\bibfnamefont{H.}~\bibnamefont{Ge}}\ and\ \bibinfo
  {author} {\bibfnamefont{D.~Q.}\ \bibnamefont{Jiang}},\ }%
  \bibfield{journal}{%
  \bibinfo {journal} {J. Stat. Phys.}\ }%
  \textbf{\bibinfo {volume} {131}},\ \bibinfo {pages} {675} (\bibinfo {year}
  {2008})\BibitemShut{NoStop}%
\bibitem{MinhAdib2009}%
  \BibitemOpen
  \bibfield{author}{%
  \bibinfo {author} {\bibfnamefont{D.~D.~L.}\ \bibnamefont{Minh}}\ and\
  \bibinfo {author} {\bibfnamefont{A.~B.}\ \bibnamefont{Adib}},\ }%
  \bibfield{journal}{%
  \bibinfo {journal} {Phys. Rev. E}\ }%
  \textbf{\bibinfo {volume} {79}},\ \bibinfo {pages} {021122} (\bibinfo {year}
  {2009})\BibitemShut{NoStop}%
\bibitem{Cover1991}%
  \BibitemOpen
  \bibfield{author}{%
  \bibinfo {author} {\bibfnamefont{T.~M.}\ \bibnamefont{Cover}}\ and\ \bibinfo
  {author} {\bibfnamefont{J.~A.}\ \bibnamefont{Thomas}},\ }%
  \emph{\bibinfo {title} {Elements of Information Theory}}\ (\bibinfo
  {publisher} {John Wiley and Sons, Inc.},\ \bibinfo {year}
  {1991})\BibitemShut{NoStop}%
\bibitem{Hummer2007}%
  \BibitemOpen
  \bibfield{author}{%
  \bibinfo {author} {\bibfnamefont{G.}~\bibnamefont{Hummer}},\ }%
  in\ \emph{\bibinfo {booktitle} {Free Energy Calculations}},\ Vol.~\bibinfo
  {volume} {86},\ \bibinfo {editor} {edited by\ \bibinfo {editor}
  {\bibfnamefont{C.}~\bibnamefont{Chipot}}\ and\ \bibinfo {editor}
  {\bibfnamefont{A.}~\bibnamefont{Pohorille}}}\ (\bibinfo {publisher}
  {Springer, Berlin},\ \bibinfo {year} {2007})\BibitemShut{NoStop}%
\bibitem{MinhAdib2008}%
  \BibitemOpen
  \bibfield{author}{%
  \bibinfo {author} {\bibfnamefont{D.~D.~L.}\ \bibnamefont{Minh}}\ and\
  \bibinfo {author} {\bibfnamefont{A.~B.}\ \bibnamefont{Adib}},\ }%
  \bibfield{journal}{%
  \bibinfo {journal} {Phys. Rev. Lett.}\ }%
  \textbf{\bibinfo {volume} {100}},\ \bibinfo {pages} {180602} (\bibinfo {year}
  {2008})\BibitemShut{NoStop}%
\bibitem{MinhChodera2009}%
  \BibitemOpen
  \bibfield{author}{%
  \bibinfo {author} {\bibfnamefont{D.~D.~L.}\ \bibnamefont{Minh}}\ and\
  \bibinfo {author} {\bibfnamefont{J.~D.}\ \bibnamefont{Chodera}},\ }%
  \bibfield{journal}{%
  \Doi{10.1063/1.3242285}{\bibinfo {journal} {J. Chem. Phys.}}\ }%
  \textbf{\bibinfo {volume} {131}},\ \bibinfo {eid} {134110} (\bibinfo {year}
  {2009})\BibitemShut{NoStop}%
\bibitem{MinhChodera2010}%
  \BibitemOpen
  \bibfield{author}{%
  \bibinfo {author} {\bibfnamefont{D.~D.~L.}\ \bibnamefont{Minh}}\ and\
  \bibinfo {author} {\bibfnamefont{J.~D.}\ \bibnamefont{Chodera}},\ }%
  \bibfield{journal}{%
  \bibinfo {journal} {J. Chem. Phys.}}%
  \emph{ in press}
   (\bibinfo {year} {2010})\BibitemShut{NoStop}%
\end{thebibliography}

%

\end{document}